\documentclass[9pt]{article}

\usepackage{latexsym}
\usepackage{amsfonts}
\usepackage{array}
\usepackage[english]{babel}
\usepackage{amsmath}
\usepackage{amssymb}
\usepackage{amsthm}
\usepackage{latexsym}
\usepackage{graphicx}

\def\bE{\begin{equation}}
\def\eE{\end{equation}}
\def\bEA{\begin{eqnarray}}
\def\eEA{\end{eqnarray}}
\def\bEAnn{\begin{eqnarray*}}
\def\eEAnn{\end{eqnarray*}}

\begin{document}
\title{{\huge \sc An elementary model of price dynamics in a financial market}\\ Distribution, Multiscaling \& Entropy}
\author{Stefan Reimann\footnote{Contact address: sreimann@iew.unizh.ch}\\
Swiss Banking Institute\\ University of Zurich}
\maketitle

\begin{abstract}
\noindent Stylized facts of empirical assets log-returns $Z$ include the existence of (semi) heavy tailed distributions $f_Z(z)$ and a non-linear spectrum of Hurst exponents $\tau(\beta)$. Empirical data considered are daily prices of 10 large indices from 01/01/1990 to 12/31/2004. We propose a
stylized model of price dynamics which is driven by expectations. The model is a multiplicative random process with a stochastic, state-dependent growth rate which establishes a negative feedback component in the price dynamics. This 0-order model implies that the distribution of log-returns is Laplacian $f_Z(z) \sim \exp(-\frac{|z|}{\alpha})$, whose single parameter $\alpha$ can be regarded as a measure for the long-time averaged liquidity in the respective market. A comparison with the (more general) Weibull distribution shows that empirical daily log returns are close to being Laplacian distributed. The spectra of Hurst exponents of both, empirical data $\tau_{emp}$ and simulated data due to our model $\tau_{theor}$, are compared. Due to the finding of non-linear Hurst spectra, the Renyi entropy (RE) $R_\beta(f_Z)$is considered.  An explicit functional form of the RE for an exponential distribution is derived. Theoretical REs of simulated asset return trails are in good agreement with the  RE estimated from empirical returns. 
\end{abstract}
\newpage
\noindent

We regard a financial market as a large complex system. Then Statistical Physics implies that (some of its) macro observables, such as prices, might be independent from its micro realizations and therefore are common to almost all realizations of the system. Indeed the statistical behavior of price fluctuations on different markets exhibit much structure that is common to 'all' financial markets. These properties are called 'stylized facts', for an enlightening survey see \cite{Cont} and also the monographs \cite{BouchaudPotters, MantegnaStanley1}. 
In turn, the existence of stylized facts suggests that price trails of different assets are realizations of a more general random/complex system, called 'a financial market'.\\

Prices are macro observables of a financial market in that their time evolution is generated by the successive trading activities of a huge amount of financial agents. This justifies to take a macroscopic perspective for modeling. The idea of our model is the following:\\

People go the financial market to make their money work. They do so by investing their money into assets. If the agent has capital $m$ to invest in asset $A$, he can buy $|A| = \frac{m}{x}$ units of this asset for its price $x$. A unit of this asset has an expected value $D$ some time later. Hence at this time, the money $m$ invested in asset $A$ has value
$$
M = \frac{D}{x} \; m
$$
If the agent is lucky, then at the expiration time $\frac{D}{x} > 1$, and his money $m$ has become valuable by a factor
$$
y \; = \frac{M}{m} =  \frac{D}{x}.
$$
It is reasonable to assume that the agent wants to spend his money in an asset of which he expects that  $\frac{D}{x} > 1$. Thus, depending on his expectation about the future value of $\frac{D}{x}$ the agents buy or sells this asset. Therefore, if the agent expects that  $\frac{D}{x} > 1$, he will buy, otherwise he will sell. This causes an increase (decrease) of demand in this asset. Due to the increase (decrease) of demand, the price will rise (fall). Therefore, price evolution is thought to depend on the expected growth rate.\\
 
Theoretical results due to our model are compared with empirical data from daily returns in the period from 01/01/1990 to 112/31/2004 of $10$ large indices listed in table \ref{indices}. Results concern i.) the non-gaussian distribution of log-returns, ii) the non-linear Hurst spectrum of return trials, and iii.) the Renyi entropy.\\

\begin{table}[h]
\begin{center}
\begin{tabular}{| c | l | l | }
\hline
1 & France & CAC 40 \\
2 & Germany & DAX 30 \\
3 & Hong Kong & {\sc Hang Seng}\\
4 & Japan & NIKKEI 225 \\
5 & Switzerland & Swiss Market Index \\
6 & Switzerland & Swiss Performance Index \\
7 & United Kingdom & FTSE 100\\
8 & United States & Dow Jones IndAvg \\
9 & United States & Nasdaq 100\\
10 & United States & S\&P500 \\
  \hline
\end{tabular}\caption{\label{indices} For each index, we considered daily data from $01/01/1990$ to $12/31/2004$ provided by {\sc Thompson Datastream}. }
\end{center}
\end{table}

\section{An expectation driven market}
The price process $X$ of an index from time $0$ to time $t$ is described a the concatenation of $n$ independent trading periods $T_\tau = [\tau,\tau+1)$, where $'[ \tau, '$ and $',\tau+1)'$ can be regarded as the opening time or the closing time of this period. The process from then is 
$$
[ \; 0 , \; t \; )  \; = \;  [ \; 0, \; 1) \; \star [ \; 1, \; 2) \; \star \hdots \; \star [ \; t-1, \; t)
$$
Let $x$ be the closing price of period $T_{\tau}$ and $x' $ the opening price of the next period. Are $x$ and $x'$ independent from each other? In reality there is night between both periods in which a lot can happen. On the other hand, reality tells us that the opening price does not differ too much from the former closing price. Now, we assume that the process can be described as a concatenation of independent processes defined on periods $T_\tau$. This means that we assume that prices $x = X_{\tau)}$ and $x' = X_{[\tau+1}$ are independent.\\

We consider a simple 1-period model $[t,t')$ of a financial market with only one asset, whose price at time $t$ is $x > 0$. Prices on this market are supposed to follow a multiplicative random process given by
\bE
x ' \; = \; \Gamma \; x
\eE
where $x'$ is the price at time $t'$. The growth rate $\Gamma > 0$ is assumed to be due to the expectations investors have at this time about future prices: Investors are supposed to build their believes about the growth rate based on some economic entity available today. A particular simple setting is that the expected growth rate is an increasing function of an expected dividend yield or earnings rate $y$, i.e.
$$
\Gamma = \Gamma^D(x) := \Phi \left(\frac{D}{x} \right),
$$ 
while $\Phi(0) = 0$. $D$ represents is a non-negative random variable which is distributed according to some (stationary) distribution $\cal F$ in some finite interval $[0,d]$. \\

In the following we assume that for all $t$, $\Phi$ is a power law, i.e.
\bE\label{scaling}
\Phi \left(\frac{D}{x} \right) \; = \; \left(\frac{D}{x} \right)^\alpha
\eE
where $\alpha \ge 0$ is a constant. The second assumption is about the distribution $\cal F$: 
\bE\label{delta}
D \sim {\cal U}(0,d).
\eE
This assumption is due to the lack of knowledge about the entity $D$. As well known, if all values within the interval are taken with equal probability, the uniform distribution is the unique distribution that minimizes information. This can be taken as the statement that all possible information are arbitraged away from a financial market. 

With
$$
D \; = \; \delta \: d, \qquad \delta \in {\cal U}(0,1)
$$
and $y(x) := \frac{d}{x}$, we finally obtain
$$
x ' \; = \; x \; \bigg( \delta \: y(x) \bigg)^\alpha
$$

\noindent
Given $x$, the probability that the gross return $ R = \frac{x'}{x}$ is larger than some $r$ is given by
\begin{eqnarray*}
F_R(r | x) &=& P[R > r | x] = P\big[ ( \delta y(x) )^\alpha > r \; | \; x\big]\\
&=& 1 - \frac{1}{y(x)} \; r^{\frac{1}{\alpha}}. 
\end{eqnarray*}
for $ 0 \le r \le \left( \frac{d}{x}\right)^\alpha$. Un-conditioning by integrating over all (initial values) $x$ implies that the unconditioned cumulative tail distribution obeys a power law
\bE\label{unconditioning}
F_R(r) = \frac{d}{2} \: r^{-\frac{1}{\alpha}}.
\eE
Therefore log-returns $Z = ln \; R$ are double exponentially distributed according to
\bE\label{z}
f_Z(z) = \frac{1}{2\: \alpha} e^{-\frac{1}{\alpha} |z|}.
\eE

\noindent
For further improvement of the model, it might be interesting to list the assumptions made. 
\begin{enumerate}
\item The financial market contains only 1 asset;
\item Due to equation \ref{unconditioning} the result holds for a 1-period model; 
\item Due to equation \ref{delta} payoffs are uniformly distributed within a fixed finite interval;
\item Due to equation \ref{scaling}, the growth rate is $\Gamma(x)$ is a power law with a constant scaling exponent $\alpha$.
\end{enumerate} 
Hence the model must be regarded as a {\sc 0-order approximations}. Particularly,  a financial market contains more than one asset, see (1), the price evolution is a process rather than can be described by a series of independent periods, see (2), (3) is reasonable only if one assumes that for all times all values in $[0,d]$ are equally probable. \\

Assumption (4) has an interpretation in terms of market liquidity. Let us assume that the volume traded is a function of the price
$$
 V = V(x) \sim \frac{1}{x}
$$
i.e. the more costly the asset is the smaller is the amount of this asset that can be bought by one unit of capital. Then the price change $Z = ln x'/x$ generated by a trade of volume $V(x)$ at price $x$ yields
\begin{eqnarray*}
Z(x) &=& \ln \frac{x'}{x} = \ln \left( \frac{D}{x}\right)^\alpha\\
&\sim& \alpha \; \ln V(x),  
\end{eqnarray*}
up to some additive constant. Therefore the price impact of a trade of size $V$ at price $x$ equals
\bE
dZ(x) \;=\; \alpha \; d  \ln V(x). 
\eE
Hence if $\alpha \approx 0$, a large trading volume is necessary to move the price.\\

Therefore $\alpha$ can be regarded as a kind of elasticity of the market at the price level $x$, that might economically understood as a measure of current market depth or its liquidity \cite{ContBouchaud}. In the light of this interpretation of the parameter $\alpha$, the fourth assumption means that liquidity should be independent of the price level and moreover constant for any time period. \\

Being aware of these assumptions, we compare our result, equation \ref{z}, with real data. Deviations of empirical return distributions from our theoretical result shall indicate which assumptions we have to modify to get a better model.  

\section{Numerical estimates of "old friends"}

Table \ref{indices} displays the set of indices considered\footnote{In this version, only the results for S\&P 500 are shown. They are in fact representative for the others considered.}. Dealing with indices, makes it necessary to discuss the meaning of the entity $D$. Originally $D$ was meant as representing the process of expected dividend pay off. Indices do not pay off dividends. Staying close to this interpretation, $D$ has to be regarded here as a sum of expected dividend flows of the constituting assets. Furthermore, expected dividend are not the only source of signals that affect expectations about the growth rate of an index. Thus signals are usually summarized as 'news' in the economic literature. Hence $D$ could also be regarded as representing the news process in the market. The broader picture therefore is that $D$ represents the process of any signal that affects expectations about growth rates of the respective market. 
\begin{figure}[h]
 \centering
 \includegraphics[width=\linewidth]{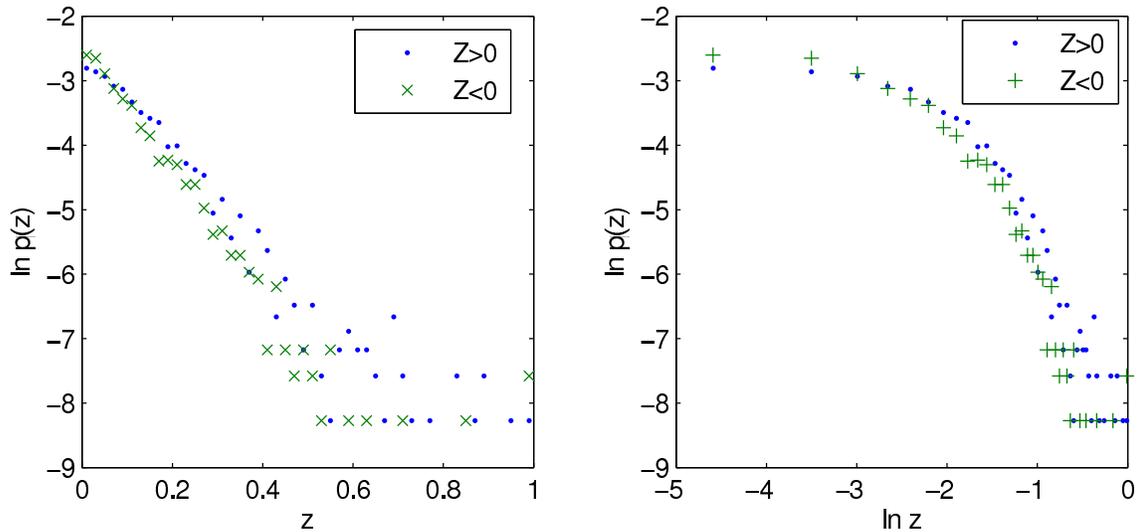}
 \caption{\label{SP500Distrs} \small Returns distributions of the {\sc S\&P 500}, left: a semi-logarithmic plot, right: a double logarithmic plot } 
\end{figure} 
The first question to be considered is to which degree our theoretical result, equation \ref{z}, agrees with empirical data. Let us first have a look at the distributions at the {\sc S\&P 500}, see Figure \ref{SP500Distrs}. The left picture is a semi-logarithmic plot and the right picture is a double logarithmic plot of daily returns from 01/01/1990 to 12/31/2004. Cross hairs mark negative returns and dots display positive returns. Recall that an exponential distribution in a semi-logarithmic plot becomes a straight line, while a straight line in a double logarithmic plot corresponds to a power-law distribution. Numerical estimates of the parameter can be found in Appendix \ref{AIC}. \\ 

\begin{figure}[h]
 \centering
 \includegraphics[width=\linewidth]{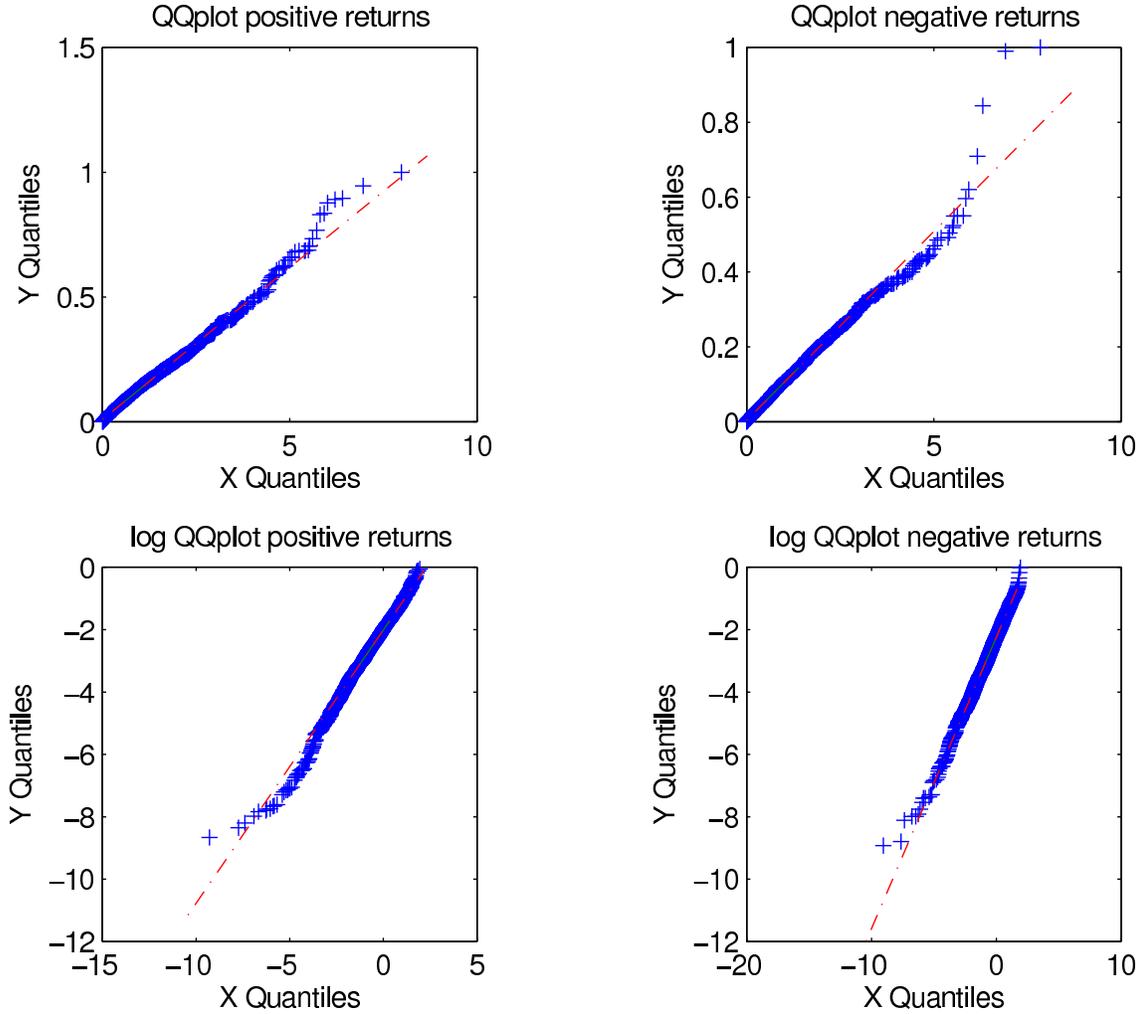}
 \caption{\label{SP500QQ} \small QQ-plots of the {\sc S\&P500} with respect to the exponential distribution}
\end{figure}

Quantile-Quantile plots provide a good descriptive method to judge about whether two sets of data come from the same distribution. A line indicates that data in both sets are very likely to come from the same distribution. As an example, let us consider the S\&P 500, see Figure \ref{SP500QQ}. \\
Figure \ref{SP500QQ} shows the QQ-plot of the empirical time series of log returns $Z$ with respect to an exponential distribution. The left column considers positive log-returns $Z_+ := Z_{\ge 0}$, while the right column is for negative log-returns $Z_ - := Z_{\le 0}$. The first row displays the QQ - plots of $Z_\pm$ which respect to the exponential, while the second row displays a QQ-plots for $\ln |Z_\pm|$ with respect to the logarithm of the exponential. The scheme see in Figure \ref{SP500QQ} thus is:
\begin{center}
{\large
\begin{tabular}{| c | c | }\hline
$Z_+$ & $Z_-$\\ \hline
$\ln Z_+$ & $\ln |Z_-|$\\ \hline
\end{tabular}
}
\end{center}
\noindent 
The QQ plots of all indices considered show the same pattern: While the middle part of each QQ-plot is linear, systematic deviations from the straight line in the QQ-plots occur for either small returns or large returns. Particularly, deviations from the diagonal for the QQ-plot of $Z_\pm$ are seen for high quantiles, while deviations form the line in the QQ-plot of $\ln |Z_\pm|$ exists for small quantiles. This means that empirical returns deviate from being exponentially distributed for very large returns, see the first row, and for very small returns, see the second row. This agrees with the observations in the respective pdf's, where we see that typically the empirical pdf has less mass in $0$ than the Laplacian, while it's tails are usually fatter than those from the Laplacian.\\ 

Fgures show the degree to which our elementary mouse model fits to empirical data: Deviations from a clear linear relation exist either for small returns and for large returns, while in the middle range, this method indicates that the proposed double exponential distribution provides a fairly good description of the empirical data. \\

\section{On multiscaling in time series of log-returns}

Stylized facts are important statistical properties since they are seen in the empirical returns of 'all' financial markets. Besides the non-Gaussian character of empirical asset returns distributions, a second important fact concerns multiscaling in the time series of empirical asset returns. The existence of multiple scales in the system implies that returns distributions are not
invariant under the choice of different time-scales, i.e. one observes that the distribution of returns with lags of the order of minutes, days, week, and
so forth deviate from each other, see \cite{SornetteCont, Gopikrishnan}.  \\

\begin{figure}[h]
 \centering
 \includegraphics[width=\linewidth]{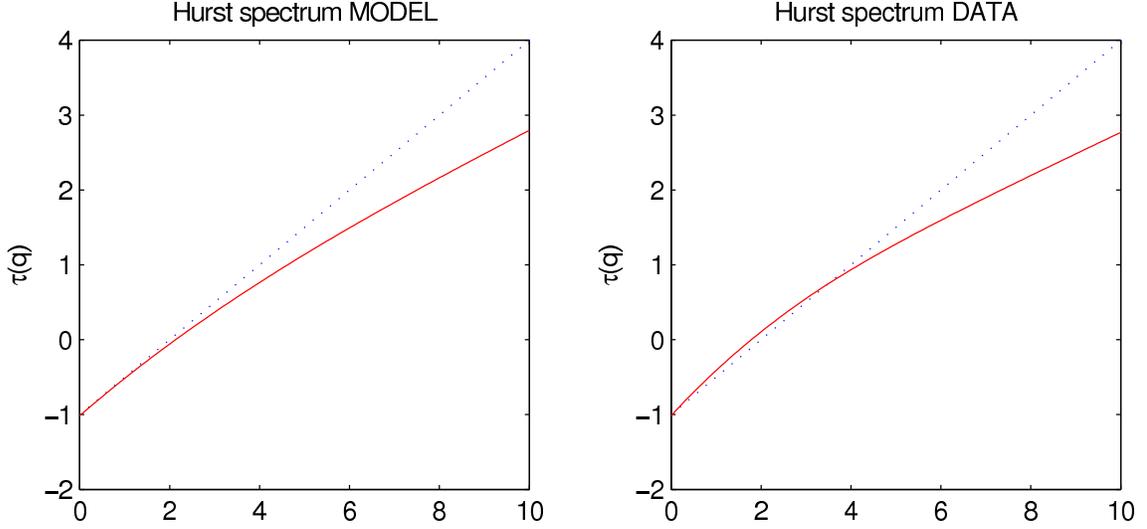}
 \vskip -0.7cm \caption{\label{mfSP500} non-linear Hurst spectrum in the time series of daily returns of the S\&P500 (right), to be compared with our model with the estimated $\alpha$, see tables \ref{negReturns}, \ref{posReturns} (left)}
\end{figure}

In the following we estimate the scaling exponent ('Hurst exponent') 
$$
{\mathbb E}\bigg( | Z(t,T)|^q \bigg) \; \sim \; t^{\tau(q)}.
$$
from time series of returns $Z(t,T) = \ln \frac{x(t+T)}{x(t)}$ with time scale $T$. In our  1-period model, $T = 1$. It can be shown that $\tau(0) = -1$, while $\tau$ is increasing and
concave. For a simple (fractional) diffusion process, $\tau$ is a linear function of $q$.   The deviation of $\tau$ from being linear, is an important issue in determining the multi scaling nature of the
underlying process. Figure \ref{mfSP500} show $\tau(q)$ as a function of $q$ for various indices compared with our model. For orientation, the dotted straight line has a slope to 1/2, which corresponds to Gaussian diffusion, while the solid line is the estimated graph of scaling exponents in our model. 
Multi-fractal processes have been proposed as a new formalism for modeling the time series of returns in finance. The major
attraction of these processes is their capability of generating various degrees of long-memory in different powers of returns - a feature that has
been found to characterize virtually all financial prices, see
\cite{Mandelbrot1, Lux, Bacry}. The prominent issue of these modeling approaches is the use of cascading processing.\\

The multifractal formalism is interesting even from another point of view. It may open the door wider to bring two disciplines closer to each other: Statistical Physics and Financial Markets, the connection being the 
R\'eny entropy'. For an introduction into the field of thermodynamics and non-linear systems and related concepts see \cite{BeckSchloegl}.  

\section{The Renyi Entropy of our stylized financial market}

\begin{figure}[h]
 \centering
 \includegraphics[width=\linewidth]{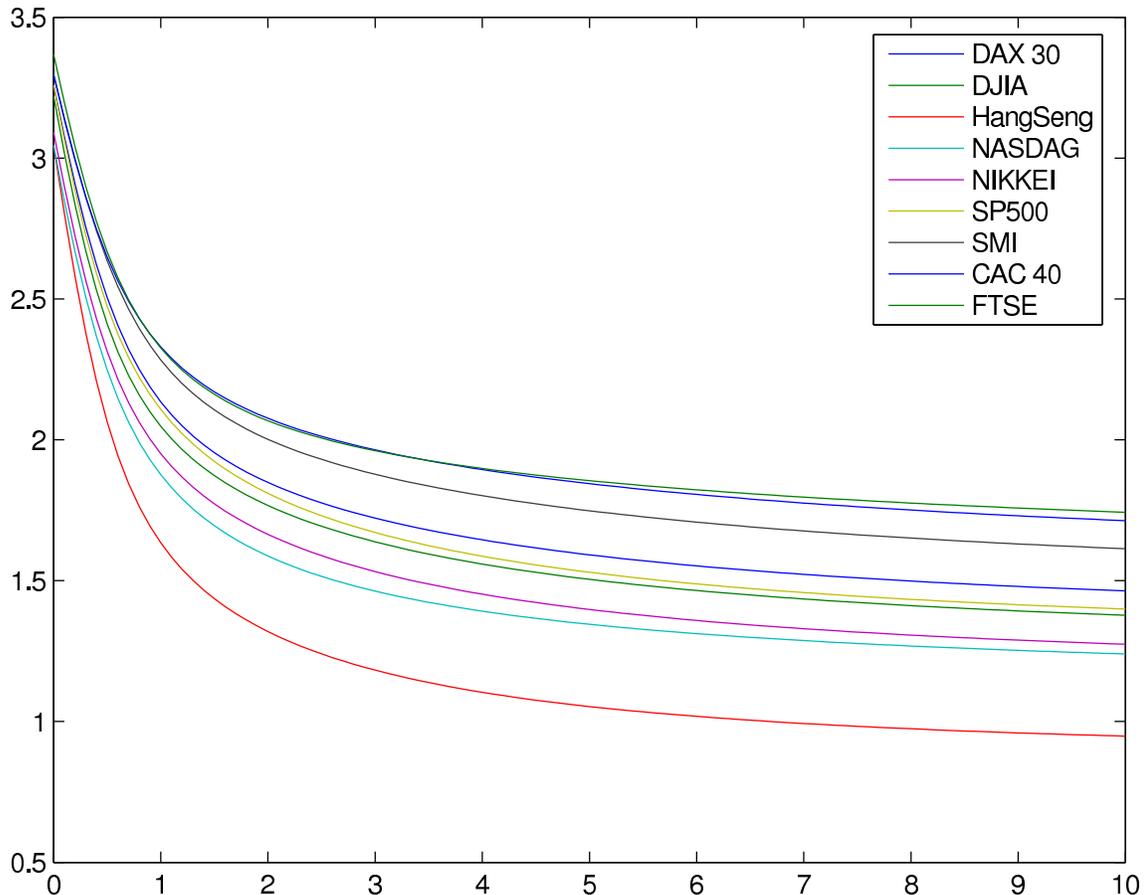}
  \caption{\label{entropies} The Renyi entropies of single indices as a function of $\beta$. }
\end{figure}

Time series of log-returns show multiscaling. The Renyi entropy has proven to be a reasonable entropy measure for multifractal systems in which long-range correlations exists. We therefore estimate the Renyi entropy defined by
\bE
R_\beta(p) \; = \; \frac{1}{1-\beta} \; \ln \sum_{i=1}^r \; p_i^\beta
\eE
for a system with $r$ micro states and a zooming parameter $\beta \in {\mathbb R}$. Figure \ref{entropies} shows the Renyi entropies for different indices. 
We will compare the theoretical Renyi entropy with the Renyi entropy found in empirical data. \\

We consider a trail of log-returns $Z$ and partition the range into $r>0$ cells each of length $1/r$. Then the k-th cells is
$$
C_k \; := \; \left[ \frac{k-1}{r} , \frac{k}{r} \right], \qquad k=1.. r.
$$
The probability $p_k$ that $Z \in C_k$ equals
$$
p_k \; = \; \int_{C_k} \; f_Z(x) \: dx
$$ 
According to equation \ref{z}, $Z$ is Laplacian distributed. Therefore we obtain
\begin{eqnarray*}
p_k&=& \frac{1}{\alpha} \int_{\frac{k-1}{r}}^\frac{k}{r} \; e^{-\frac{x}{\alpha}} \; dx\\
&=& e^{-\frac{k}{\alpha\: r}} \; \left( e^{\frac{1}{\alpha \: r}} - 1\right)
\end{eqnarray*}
Summing probabilities over the sample space $\{1,..,r\}$ gives
\begin{eqnarray*}
\sum_k p_k &=& \frac{1}{\alpha} \sum_{i=1}^r \int_{\frac{k-1}{r}}^\frac{k}{r} \; e^{-\frac{x}{\alpha}} \; dx \; = \; \frac{1}{\alpha} \int_0^1 \; e^{-\frac{x}{\alpha}} dx\\
&=& 1 - e^{-\frac{1}{\alpha}}.
\end{eqnarray*}
For later purposes we define
\bE
C_\beta \; := \; \frac{e^{-\frac{\beta}{\alpha \: r}} - 1}{1 - e^{-\frac{\beta}{\alpha}} }
\eE
Therefore we normalize probabilities by
$$
\pi_k \; = \; C_1 \; e^{-\frac{k}{\alpha \: r}} 
$$
The Renyi entropy yields
\begin{eqnarray*}
R_\beta(p) &=& \frac{1}{1-\beta} \; \ln \sum_k \pi_k^\beta \; = \; \frac{1}{1-\beta} \; \ln \sum_k  C_1^\beta 
e^{-\frac{\beta \: k}{\alpha \: r}}\\
&=& \frac{1}{1-\beta} \left[ \ln C_1^\beta  + \ln \sum_k e^{-\frac{\beta \: k}{\alpha \: r}}\right]
\end{eqnarray*}
From $\sum_{k=1}^r \: e^{-\frac{\beta \: k}{\alpha \: r}} =  \frac{1 - e^{-\frac{\beta}{\alpha}} }{e^{-\frac{\beta}{\alpha \: r}} - 1} = \frac{1}{C_\beta}$, we obtain for the Renyi entropy
\bE\label{R}
 R_\beta(p) \; = \;  \frac{1}{1-\beta} \;  \ln \; \frac{C_1^\beta}{C_\beta}
\eE
To estimate the parameter $\alpha$ from measuring the Renyi entropy, it is therefore sufficient to consider two cases, $\beta = 0,1$, where $R_0(p) := \lim_{\beta \to 0} R_\beta(p)$ and $R_1(p) := \lim_{\beta \to 1} R_\beta(p)$. From its definition it immediately follows the well known relation
\bE\label{r}
R_0(p) \; = \; \ln r
\eE
Further, in the limit $\beta \to 1$, the Renyi entropy becomes the Shannon-Boltzmann entropy: $R_1(p) = - \sum_k \pi_l \ln \pi_k$ and therefore
\begin{eqnarray*}
R_1(p) &=& -\sum_k C_1 e^{-\frac{k}{\alpha \: r}} \; \ln \left( C_1 e^{-\frac{k}{\alpha \: r}}\right)\\
&=& - C_1 \sum_k e^{-\frac{k}{\alpha \: r}} \left( \ln C_1 - \frac{k}{\alpha \: r}\right)\\
&=& -\ln C_1 + \frac{C_1}{\alpha \: r} \sum_k \: k \; = \; -\ln C_1 + \frac{C_1}{\alpha \: r} { r \choose 2}
 \end{eqnarray*}
 so that we obtain
 \bE\label{alpha}
 R_1(p) \; = \; -\ln C_1 + \frac{r+1}{2\: \alpha} \; C_1.
 \eE
 Since $C_1 = C_1(\alpha,r)$, equations \ref{r} and \ref{alpha} allow to estimate the distribution parameter $\alpha$ from empirical data - under the hypothesis that the distribution of $Z$ is exponential.\\
 
\begin{figure}[h]
 \centering
 \includegraphics[width=\linewidth]{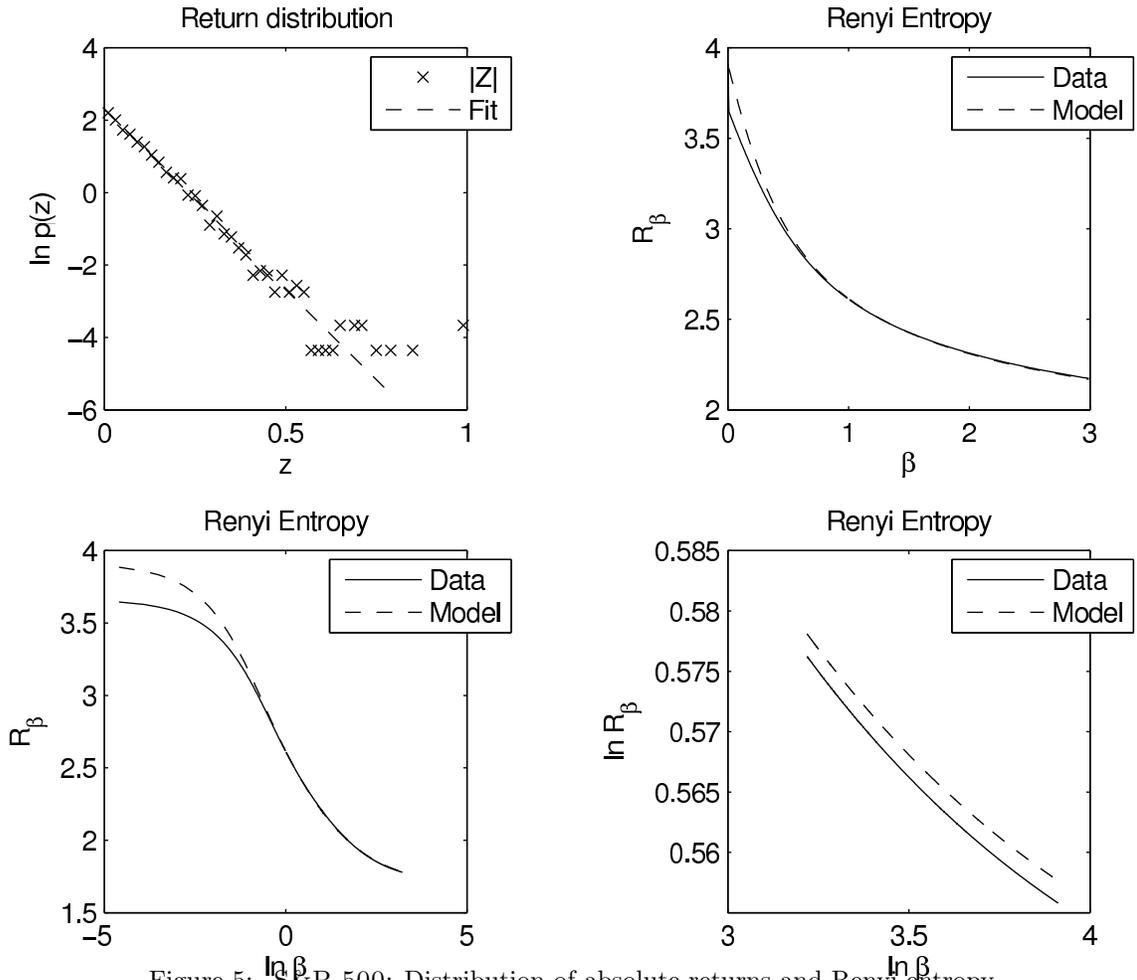}
 \vskip -0.7cm \caption{\label{SP500Entropy} {\sc S\&P 500}: Distribution of absolute returns and Renyi entropy}
\end{figure}

We calculate the Renyi entropy for return trails of each single index $1, .., 10$, see Table \ref{indices}, for a fixed number of cells $r=30$, see Figure \ref{SP500Entropy}. Former estimates of the asymmetry of the distribution of negative returns and positive returns, measured by the parameters $\alpha_\pm$ showed that the distribution is (almost) symmetric. Hence we considered the trail of absolute returns, i.e. $|\ln Z|$. We normalized returns to the unit interval by considering $|Z^*| = \frac{|Z|_{emp}}{\max |Z|_{emp}}$. The left upper picture shows the distribution of $|Z^*|$ and a fit with respect to the exponential distribution given the (adjusted) $\alpha$ estimated from the exponential distribution as in Appendix \ref{AIC}. The upper right picture shows the resulting empirical Renyi entropy (solid line) and the graph of $R_\beta(p)$ from equation \ref{R} (dashed line). 
The prediction of our model agrees fairly well with the data. The logarithmic plots in the lower row show the deviations for small $\beta \in [0,3]$ and large $\beta \in [25,50]$, see Figure \ref{SP500Entropy}. Recalling that $\beta$ is a zooming factor, which places emphasis on different probability regimes, these deviations agree with the findings stated earlier: Our model gives too much mass to high probabilities, related to small returns, while it does not give enough mass in the tails, i.e. for large returns. 

 \section{Conclusion}
\begin{quotation}
{\flushright
{\em  All models are wrong but some are useful.}

-- G. E. P. Box, 1979 

}
\end{quotation}
Our aim is to understand price dynamics on a financial market. The existence of stylized facts suggests that price trails of different financial markets might be regarded as different realizations of a more general stochastic system, called 'The financial market'. If so then the question is about the nature of this system. Since prices are macro-observables of a financial market, the model about price dynamics is defined on the macro level. Due to the set of assumptions used in its derivation, this model is an approximation in itself. Results therefore can also only as zero-approximations as indicated by the list of assumptions made. We estimated three major properties: the distribution of (logarithmic) asset returns, the Hurst exponent in their time series, and finally the R\`eny entropy. Although the model is a zero-order approximation, theoretical results are already in fairly good agreement with real data. First-order corrections concerning the set of assumptions made should improve these theoretical findings. Taking all this together, this model might serve as a good starting point for further improvements. These steps should follow from observing where our theoretical results deviate from empirical data and by successively and modestly modifying the assumptions made. 

\section*{Acknowledgment}

The author thanks  V. B\"ohm and U. Horst for motivating discussions as well as Urs Schweri and V. V. d'Silva for their valuable cooperation.

\newpage
\begin{appendix}
\section{Quantitative estimation of parameters}\label{AIC}

We compare two distributions with respect to their goodness to fit empirical data. One is the exponential with parameter $\alpha$, while the other one is the Weibull with parameters $(a,\mu)$
\begin{eqnarray*}
f^{exp}_Z(z) &=& \frac{1}{\alpha} \: e^{-\frac{z}{\alpha}}\\
f^{WB}_Z(z) &=& \frac{m}{z}
 \left( \frac{z}{a} \right)^{m} \; e^{\left( -\frac{z}{a}\right)^m}
 \end{eqnarray*}
The exponential distribution is obtains from the Weibull in the case $\mu=1$, while in the limit $\mu \to 0$, $f_Z(z)$ approximates the Pareto distribution arbitrarily well, see \cite{Malevergne}. 
Fits of the distributions for positive and negative returns due to these distributions therefore give respective parameters
\begin{eqnarray*}
{\mathbf \alpha} &=& \left( \alpha^+, \alpha^-\right)\\
{\mathbf \beta} &=& \left( a^+,m^+ \; ; \; a^-, m^-\right).
\end{eqnarray*}
To make differences more obvious, we standardize log returns in the usual way
$$
Z'_\pm  \; = \; \frac{Z_\pm  - \langle Z_\pm  \rangle}{\sigma(Z_\pm)},
$$
where $Z'_\pm$ are the positive and negative returns, respectively, and $\sigma$ denotes the respective standard deviation.  Therefore the estimates parameter $\mu$ is of special interest. In the following we summarize log likely fits of positive and negative returns, respectively, to the exponential distribution and the Weibull distribution, giving parameters $\alpha^\pm$ and $(a^\pm, \mu^\pm)$ respectively.\\

Distributions are close to being symmetric $\alpha_+ \approx \alpha_-$, see tables \ref{posReturns} and \ref{negReturns}. Furthermore $ 0 \ll m^\pm \approx 1$, i.e. their distributions are close to an exponential distribution. This finding is also supported by considering the respective entropic distances between the empirical distribution and the exponential and Weibull distribution respectively as formalized by the Akaike's Information Criterion
$$
AIC = - 2 \; \log \bigg( {\cal L}(\hat{\theta} | x)\bigg) + 2\: K,
$$
where $K$ is the number of parameters, here K=1 for the exponential and $K=2$ for the Weibull distribution. The term $2 \:K$ can be considered as a penalty for introducing additional parameters. 
${\cal L}(\hat{\theta} | x)$ is the maximum log-likelihood of the parameter $\theta$ given the data $x$. AIC is in particular useful for nested model such as the exponential and the Weibull in this case. However, we considered the ratio of the AIC's of both models
$$
\mbox{$ratio$} \; = \;  \frac{{\cal L}(exponential)+1}{{\cal L}(Weibull) + 2},
$$
As seen in tables \ref{posReturns} and \ref{negReturns}, the ratio is close to $2/3$, which indicates that the approximation that the distribution of log returns is an exponential is quite well and that we can trust the estimated parameters $\alpha_k$.

\begin{table}[p]
\begin{center}
\begin{tabular}{| l || l | ccc | c |}\hline
INDEX	 &	\# $Z<0$&		$\alpha^-$ &	    $a^-$ & $\mu^-$ & $ratio$ \\ \hline
DAX 30	& 1806 & 0.6985     & 0.6613     & 0.8885     & 0.6671\\
SWISS SMI EXP	& 1066 & 0.7021     & 0.6762     & 0.9209     & 0.6673\\
FRANCE CAC 40 & 1849 & 0.7054     & 0.6869     & 0.9399     & 0.6671\\
FTSE 100	& 1840 & 0.7074     & 0.6997     & 0.9742     & 0.6671\\
SWISS SPI EXTRA	&	955  & 0.7342     & 0.6891     & 0.8820     & 0.6673 \\
DOW JONES   & 1793	& 0.6891     & 0.6713     & 0.9437     & 0.6671\\
Hang Seng &	1808	& 0.6393     & 0.6000     & 0.8815     & 0.6671\\
NASDAQ 100 &1745& 0.7077     & 0.6887     & 0.9405     & 0.6671\\
NIKKEI 500 &	1864& 0.7500     & 0.7698    & 1.0688     & 0.6671\\
S\&P 500 	&	1799		 & 0.6869     & 0.6663     & 0.9353     & 0.6671\\
  \hline
\end{tabular}
\caption{\label{negReturns} Parameter estimation for negative returns of the indices considered}
\end{center}
\end{table}

\begin{table}[p]
\begin{center}
\begin{tabular}{ | l || l | ccc | c |}\hline
INDEX	 & \# $Z>0$	&	$\alpha^+$ &	$a^+$& $\mu^+$  &  $ratio$\\ \hline
DAX 30	&1964& 0.7057     & 0.7280     & 1.0840     & 0.6671 \\
SWISS SMI EXP	&1190 & 0.7003     & 0.7186     & 1.0683     & 0.6673\\
FRANCE CAC 40 &	1925 & 0.7454     & 0.7786     & 1.1285     & 0.6671\\
FTSE 100	& 1948& 0.7339     & 0.7638     & 1.1129     & 0.6671\\
SWISS SPI EXTRA	&1283 & 0.6634     & 0.6984     & 1.1615     & 0.6673 \\
DOW JONES  & 1979	& 0.7223     & 0.7478     & 1.0948     & 0.6671\\
Hang Seng&	1904    & 0.7038     & 0.7105     & 1.0229     & 0.6671\\
NASDAQ 100 & 2033  & 0.6933     & 0.7070     & 1.0485     & 0.6671\\
NIKKEI 500&	1825    & 0.6592     & 0.6173     & 0.8699     & 0.6670\\
S\&P 500 	&1983  & 0.7207     & 0.7373      &1.0589      & 0.6671\\  \hline
\end{tabular}
\caption{\label{posReturns} Parameter estimation for positive returns of the indices considered}
\end{center}
\end{table}
\end{appendix}
\end{document}